\input{epsfig.sty}
\documentstyle[preprint,aps]{revtex}
\begin{document}
\draft
\title{Induction of non--d-wave order-parameter components by currents
in d-wave superconductors}
\author{Martin Zapotocky\rlap,$^{(a)}$\cite{MZ}
Dmitrii L.~Maslov\rlap,$^{(a,b)}$\cite{DLM} and
Paul M.~Goldbart$^{(a)}$\cite{PMG}}
\address{
$^{(a)}$Department of Physics and Materials Research Laboratory,\\
University of Illinois at Urbana-Champaign, Urbana, Illinois 61801, USA\\
$^{(b)}$Institute for Microelectronics Technology, 
Academy of Sciences of Russia,\\
Chernogolovka, 142432 Russia}
\date{\today}
\maketitle
\begin{abstract}
It is shown, within the framework of the Ginzburg-Landau theory
for a superconductor with {$d_{x^2-y^2}$}\  symmetry, that the passing of a
supercurrent through the sample results, in general,
in the induction of order-parameter components of distinct symmetry. 
The induction of $s$-wave and
{$d_{xy(x^2-y^2)}$}-wave components are considered in detail. 
It is shown that in both cases the order parameter remains gapless;
however, the structure of the lines of nodes and the lobes of the 
order parameter are modified in distinct ways, and the magnitudes of
these modifications differ in their dependence on the ($a$-$b$ plane)
current direction.
The magnitude of the induced $s$-wave component is estimated using the
results of the calculations of Ren et al.~[Phys.~Rev.~Lett. {\bf 74},
3680 (1995)], which are based on a
microscopic approach.
\end{abstract}
\narrowtext

\section{Introduction}

As a result of recent experimental~\cite{REF:Dale,REF:Bonn,REF:Moler}
and theoretical~\cite{REF:theory?} work there is an emerging consensus
that the symmetry of the superconducting state in the high-temperature
superconducting materials is that of $d_{x^2-y^2}$-wave.  Given this
situation, it seems worthwhile to explore the phenomenological
implications of such a state, even if the microscopic origin of the
superconductivity has not yet been fully established. In particular,
the following feature of the phenomenological theory of $d$-wave
superconductors has attracted the attention of a number of workers.
In the absence of any external agents (e.g., magnetic fields,
surfaces, currents, etc.), 
the only component of the superconducting order parameter that has a
non-zero mean equilibrium value is the component with {$d_{x^2-y^2}$}\ 
symmetry,
 the other (i.e. subsidiary) components exhibiting equilibrium
fluctuations
around a mean value of zero.  External forces can give rise to non-zero mean 
values of
subsidiary components. Interestingly, general symmetry
considerations permit a coupling between the gradients of {$d_{x^2-y^2}$}\ 
and of
other components~\cite{REF:Joynt}. In particular, this means that {\it
any\/} inhomogeneity in {$d_{x^2-y^2}$}\ acts as a {\it source\/} of
inhomogeneities in other components and, therefore, as a source of
these components themselves. This mechanism has been exploited by a
number of authors.  Notably, Volovik~\cite{REF:Volovik}, followed by
Soininen et al.~\cite{REF:Soin_PRL} and other 
workers~\cite{REF:Ren,REF:Soin_Review},
have predicted that the vortices in a {$d$}-wave superconductor should
exhibit a rich structure, in which $s$- and $d$-components of the
order parameter co-exist. Furthermore, the surface regions of these
superconductors are predicted to be in 
a mixed $s$-$d$--state~\cite{REF:Sauls}. 

In this work we pursue one further consequence of the gradient
coupling mechanism, viz., we predict that an external current can
induce non-zero subsidiary components of the superconducting order
parameter via the (current-induced) inhomogeneity of the dominant 
(i.e.\ {$d_{x^2-y^2}$}) component. As opposed to the cases of 
vortices~\cite{REF:Volovik,REF:Soin_PRL,REF:Ren,REF:Soin_Review} and 
surfaces~\cite{REF:Sauls}, which both have an {\it amplitude\/} 
variation of the dominant ({$d_{x^2-y^2}$}) component,
the induction of subsidiary components by the current requires only
its {\it phase} variation.  In Section~II, we present the Ginzburg-Landau
theory of the current-induced $s$-component in a $d$-wave
superconductor. Our treatment is based on the Ginzburg-Landau (GL)
theory for a {$d$}-wave superconductor~\cite{REF:Joynt,REF:Soin_Review}
that incorporates the effects of $s/d$-coupling.  As it is not yet
clear which of the subsidiary components has the strongest coupling to
the {$d_{x^2-y^2}$}\ -component, we then (in Section III) extend this 
treatment to include
subsidiary irreducible representations of the tetragonal ($D_4$)
symmetry group other than $s$-wave. Finally, we discuss some
experimental settings in which current-induced subsidiary components
of the order parameter might be observable.

\section{The case of 
$\lowercase{s}/\lowercase{d_{x^2-y^2}}$\ --coupling}

First we focus on the case of $s$/{$d_{x^2-y^2}$}\ -coupling, 
in which the order parameter has two
spatially varying complex components $s({\bf r})$ and $d({\bf r})$. We neglect 
the magnetic fields induced by the current~\cite{NOTE:field}. 
The GL equations for $d$- and $s$-components
were derived in Refs.~\cite{REF:Joynt,REF:Soin_Review}; they are
\begin{mathletters}
\begin{eqnarray}
(-{\gamma_{d}}\nabla^2+{\alpha_{d}})d+{\gamma_{v}}(\nabla_x^2-\nabla_y^2)s+
2\beta_2|d|^2d+\beta_3|s|^2d
+2\beta_4s^2d^{*}&=&0,\label{EQ:GLd}\\
(-{\gamma_{s}}\nabla^2+{\alpha_{s}})s+{\gamma_{v}}(\nabla_x^2-\nabla_y^2)d+
2\beta_1|s|^2s+\beta_3|d|^2s
+2\beta_4d^2s^{*}&=&0,\label{EQ:GLs}
\end{eqnarray}
\end{mathletters}
where $\gamma_{\rho}\equiv{\hbar}^2/2m_{\rho}$, and $\rho=d,s,v$.
The current density is given by
\begin{eqnarray}
{\bf J}&=&\frac{e\hbar}{i{m_{d}}}\big\{d^{*}\nabla d-\mbox{c.c.}\big\}+
\frac{e\hbar}{i{m_{s}}}\big\{s^{*}\nabla s-\mbox{c.c.}\big\}
\nonumber\\
&-&{\hat x}\frac{e\hbar}{i{m_{v}}}\big\{s^{*}\nabla_xd-d\nabla_xs^{*}-
\mbox{c.c.}
\big\}
+{\hat y}\frac{e\hbar}{i{m_{v}}}\big\{s^{*}\nabla_yd-d\nabla_ys^{*}-
\mbox{c.c.}\big\}\label{EQ:J},
\end{eqnarray}
where we have chosen the effective charge $e$ to be twice the electron charge. 
The parameters of the GL equations 
(\ref{EQ:GLd},\ref{EQ:GLs}) are chosen in such a 
way~\cite{REF:Soin_PRL,REF:Soin_Review} that, in the absence 
of the current, $|d|>0$ and $s=0$. In the presence of the
current, we assume that $s$ is nonzero but small compared with $d$
and, therefore, can be analyzed perturbatively. In this way,
the inhomogeneity in $d$ acts as a source for $s$. To the zeroth order in
perturbation theory, we have
\begin{equation}
d=d_0\,{\rm e}^{i{\bf q}_0 \cdot {\bf r}},\quad\quad s=s_0=0.
\label{EQ:pt0}
\end{equation}
The
amplitude $d_0$ and the wave vector ${\bf q}_0$ are found in the usual
way~\cite{REF:deGennes} from Eqs.~(\ref{EQ:GLd},\ref{EQ:J}) with $s$
having been set to zero:
\begin{mathletters}
\begin{eqnarray}
q_0&=&\xi_{d}^{-1}(1-f^2)^{1/2},\ {\rm and} \quad{\bf q}_0\parallel{\bf J}\
label{EQ:q},\\
j&\equiv & J/J_{c}=3\sqrt{3}f^2(1-f^2)^{1/2}/2,\label{EQ:j}
\end{eqnarray}
\end{mathletters}where $f=d_0 / \sqrt{|{\alpha_{d}}| / 2 \beta_2}$ is 
the dimensionless {$d$}-wave order 
parameter normalized by its equilibrium value,
 $\xi_{d}\equiv \sqrt{\hbar^2/2{m_{d}} |{\alpha_{d}}|}$ is the
correlation length of the {$d$}-wave order parameter, and
$J_c$ is the critical current density. The dependence of $f$, and thus 
of $d_0$, on $j$ is given by the implicit relation (\ref{EQ:j}). In 
particular, $f=1$ for $j=0$, and $f$ approaches the value of 
$\sqrt{2/3}$ from above as $j$ approaches $1$ from below; for $j >
1$ we have  
$f=0$. To 
first order in perturbation theory, $s$ acquires a non-zero value and
$d$ changes from its zeroth-order value. This also leads to a change
in the right-hand side of Eq.~(\ref{EQ:J}), which determines the wave vector
of the order parameter for a given current density.
This means that at this order the wave vector found in the
previous order is changed. Thus the appropriate {\it Ansatz\/} at first order
is
\begin{mathletters}
\begin{eqnarray}
d&=&d_0{\rm e}^{i({\bf q}_0+{\bf q}_1)\cdot{\bf r}}+d_1{\rm e}^{i({\bf q}_0+
{\bf q}_1)\cdot{\bf 
r}},\\
\label{EQ:pt1d}
s&=&s_1{\rm e}^{i({\bf q}_0+{\bf q}_1)\cdot{\bf r}},
\label{EQ:pt1s}
\end{eqnarray}
\end{mathletters}where quantities with the subscript $1$ are small compared 
to those with
the subscript $0$. As can be readily checked, this {\it Ansatz} satisfies
Eqs.~(\ref{EQ:GLd},\ref{EQ:GLs},\ref{EQ:J}). Keeping only
terms linear in $d_1$, $s_1$ and ${\bf q}_1$, and after some lengthy
but straightforward algebra~\cite{REF:Anote}, we obtain:
\begin{equation}
\frac{s_1}{d_0}=\frac{{\gamma_{v}} q_0^2 \cos(2\phi)}
{{\gamma_{s}} q^2_0+{\alpha_{s}}-\frac{9{\gamma_{v}}^2 q_0^2
\cos^2(2\phi)}{-3{\gamma_{d}} q_0^2+{\alpha_{d}}+6\beta_2d_0^2}
-4\frac{{\gamma_{v}}^2}{{\gamma_{d}}}q_0^2+(\beta_3+2\beta_4)d_0^2},
\label{EQ:GLs1}
\end{equation}
where $\phi$ is the angle between ${\bf J}$ and the 
$x$-axis in the $a$-$b$--plane. Already at 
this stage two conclusions can be made. First, the amplitude of
the induced $s$-component depends not only on the {\it amplitude\/} but 
also on the
{\it direction\/} of the current: $|s|$ is maximal for a current flowing 
along the major crystallographic axes (i.e., for
$\phi=0$ or $\phi=\pi/2$) and is zero
(at this order of perturbation theory) for a current flowing along the
diagonal of the unit cell (i.e., for $\phi=\pi/4$). Second, the rather 
cumbersome expression (\ref{EQ:GLs1}) is simplified considerably   
for temperatures $T$ very close to the critical temperature 
$T_{\rm c}$ (i.e., in the critical regime, when the GL approach is strictly 
valid).
In the limit $T\to T_{\rm c}$ all the terms in the denominator of 
Eq.~(\ref{EQ:GLs1}) 
that contain $q_0$ become small because $\xi_{d}$ diverges, and the
last term in the denominator is small because $d_0$ is small. On the other 
hand,
as $T_{\rm c}$ is not a critical temperature for the $s$-wave component, 
${\alpha_{s}}$ is nonzero in this limit; thus ${\alpha_{s}}$ dominates the 
denominator.
Equation~(\ref{EQ:GLs1}) then takes on the simpler form:
\begin{equation}
\frac{s_1}{d_0}=
\frac{{m_{d}}}{{m_{v}}}
\frac{|{\alpha_{d}}|}{{\alpha_{s}}}
(1-f^2)\cos(2\phi).
\label{EQ:GLs1simp}
\end{equation}
Equation~(\ref{EQ:GLs1simp}) shows that for generic values of the parameters,
i.e., for ${m_{d}}\simeq{m_{v}}$ and $\phi\simeq 1$, the
smallness of $s_1$ with respect to $d_0$, and thus  the validity of the
perturbation theory, is guaranteed by the smallness of the ratio
$|{\alpha_{d}}|/{\alpha_{s}}\propto(T_{\rm c}-T)$.  Therefore, in the 
critical region, 
the perturbation theory is valid even for currents that are not small
compared to $J_{\rm c}$ (i.e., for values of $f$ that are not close to $1$). 

\begin{figure}
\epsfig{figure=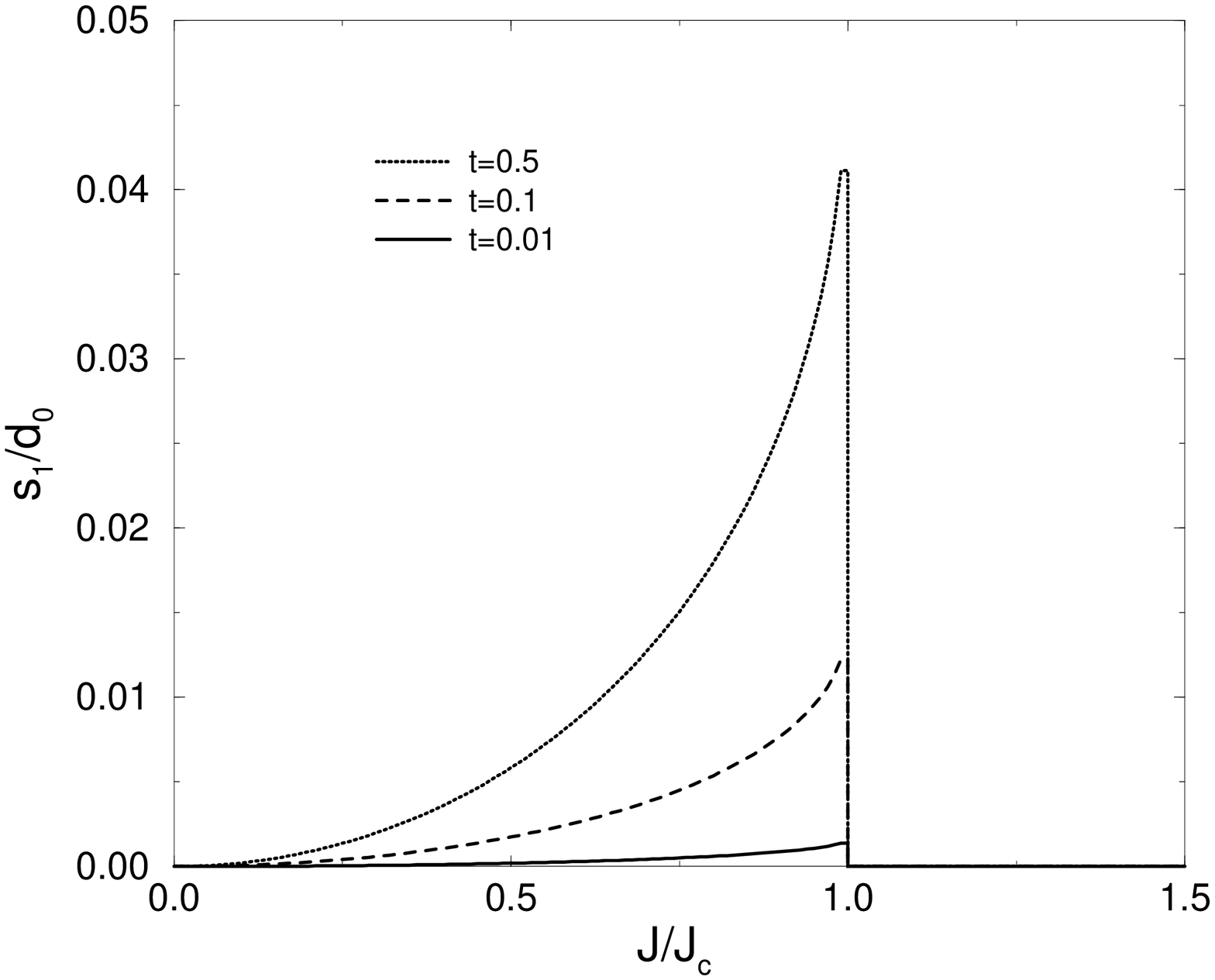, height=12cm,rheight=14cm,angle=270}
\caption{The magnitude $s_1$ of the s-wave component induced by the
applied current density $J$, for the reduced temperature values 
$t=0.5$ (dotted line), $t=0.1$ (dashed line), and $t=0.01$ (solid
line). The curves are obtained from Eqs.~(4b) and (9) with $\phi=0$.
\label{FIG:s_vs_j}}
\end{figure}

In order to obtain (semi-)quantitative estimates for the amplitude of
the induced $s$-wave order parameter as given by Eq.~(\ref{EQ:GLs1}),
we need to know the values of the phenomenological parameters of the GL
theory. These can be estimated, e.g., by comparing Eqs.~(\ref{EQ:GLd},
\ref{EQ:GLs},\ref{EQ:J}) with the GL equations that were derived
recently by Ren et al.~\cite{REF:Ren} from the Gor'kov equations for a 
particular microscopic model of pairing interactions~\cite{REF:Feder}. 
Reference~\cite{REF:Ren} gives the following ratios of the
phenomenological GL parameters:
\begin{mathletters}
\begin{eqnarray}
{m_{s}}:{m_{d}}:{m_{v}}&=&1:2:2,\\
\beta_1:\beta_2:\beta_3:\beta_4&=&1:3/8:2:1/2,\\ 
|{\alpha_{d}}|/{\alpha_{s}}&=&\lambda_{d}\ln(T_{\rm c}/T)/2(1+V_s/V_d),
\label{EQ:params}
\end{eqnarray}
\end{mathletters}
 where $\lambda_{d}$ is the BCS coupling
constant in the {$d$}-wave channel, and $V_{d}$ and $V_{s}$ are interaction 
parameters,
which in the model of Ref.~\cite{REF:Ren} describe nearest-neighbor
attraction and on-site repulsion, respectively. As mentioned 
in Ref.~\cite{REF:Ren}, the $s$-wave component is induced by
inhomogeneities in the {$d$}-wave component even if $V_{s}=0$.
For lack of better information about $V_{s}$, we set it to zero, which
does not 
significantly affect our results. By using the ratios of the GL parameters
given above, 
Eq.~(\ref{EQ:GLs1}) is cast into the following form:
\begin{equation}
\frac{s_1}{d_0}=\frac{1}{2}\frac{(1-f^2) \cos (2\phi)}
{\frac{1}{\lambda_d\ln(T_{\rm c}/T)}-\frac{9}{4} \frac{(1-f^2)^2 \cos^2(2\phi)}
 {3 f^2 - 1} +1+f^2}.
\label{EQ:s1micro}
\end{equation}
To estimate the BCS coupling constant $\lambda_{d}$, 
we use the result of Monthoux and Pines~\cite{REF:Month}, 
who find that, in a spin-fluctuation model,
the value of $T_{\rm c}=90$\,K is obtained for $\lambda_{d}$ close to $1$
(the precise value depending on the doping).
Solely for illustrative purposes, we 
use the value $\lambda_{d}=1$. 
The dependence of $s_1/d_0$ on $j$ is shown in Fig.~\ref{FIG:s_vs_j}
for three values of $t\equiv (T_{\rm c}-T)/T_{\rm c}$. 
For the values $t=0.01$ and $0.1$ the result given
by  Eq.~(\ref{EQ:s1micro}) is very close to that given by its
simplified version Eq.~(\ref{EQ:GLs1simp}). For $t=0.5$ the result
given by Eq.~(\ref{EQ:s1micro}) is approximately one half of that given
by  Eq.~(\ref{EQ:GLs1simp}). 
At temperatures far below $T_{\rm c}$ (i.e. for $t \simeq 1$), the
microscopic derivation of the GL parameters leading to
Eq.~(\ref{EQ:s1micro}) is not strictly valid, and the term $\ln(T_{\rm c}/T)$ 
is
expected to be replaced by $t$ (note that $\ln(T_{\rm c}/T)=t$ for $t \ll
1$) , thus avoiding the apparent singular behavior in Eq.~(\ref{EQ:s1micro}).

The presence of the $s$-wave component, which according to
Eqs.~(\ref{EQ:pt1d},\ref{EQ:pt1s}) is in-phase with the {$d$}-wave component,
implies that excitations with momentum along the $\Phi=\pm\pi/4$
directions are no longer gapless. Rather, they have the energy gap given by
$\Delta_{s}=\Delta_{d} s_1/d_0$, where $\Delta_{d}$ is the maximum
value of the {$d$}-wave gap in the absence of the currents. 
The lines of nodes, oriented along the $\Phi=\pm\pi/4$ directions in the
absence of current, are now rotated by the angle 
$\delta\Phi={1\over 2}\cos^{-1}(\Delta_{s}/\Delta_{d})$
[see Figs.~(\ref{FIG:d+s+a}a,b)].
The $k$-space structure of the order
parameter undergoes an {\it orthorhombic distortion\/}, i.e., the 
current-induced s-wave component mimics the effect of having an
orthorhombic (rather than tetragonal) lattice and no supercurrent. 
By using the typical
value of $\Delta_{d}\simeq 100$\,K, we see, e.g.,
that for $t=0.5$ (i.e., for $T=0.5T_{\rm c}$) and for $j=0.5$ the
 gap $\Delta_{s}\simeq 1$\,K, and the lines of nodes rotate by
$\delta\Phi \simeq 0.3^o$.  We must emphasize that at lower temperatures
(i.e., when $t\simeq 1$) and for currents comparable to
 the critical current, there is no longer a natural small parameter
in the theory that would 
automatically guarantee the smallness of $s_1$ with
respect to $d_0$. The fact that $s_1$ remains small
even in this region is due to the particular choice of the ratios
of the GL parameters. However, the GL theory is not expected
to be quantitatively correct in this region, so the microscopic theory
might give other numerical values of $\Delta_s$ and $\delta\Phi$. The absence
of a natural small parameter suggests that these values
might be larger than those given by the perturbative treatment of the
GL equations.

\section{Coupling to other subsidiary order-parameter components}

So far, we have considered the coupling of the dominant 
{$d_{x^2-y^2}$}-component to the
(subsidiary) $s$-wave component, which is taken as the main
subsidiary component in the microscopic approaches 
of~\cite{REF:Soin_PRL,REF:Ren}. In general, the GL theory should incorporate
all irreducible representations of the $D_4$ symmetry group; it
is then the task of a microscopic theory to determine the dominant, and
leading subdominant, components. Although the growing consensus is that
the leading component corresponds to the {$d_{x^2-y^2}$}\ representation, 
it is not
clear, at present, which representation describes the subleading 
component~\cite{REF:Sauls}. Therefore, we now extend the treatment presented
above to include couplings between the {$d_{x^2-y^2}$}\ component and 
components
of the order parameter other than $s$-wave. 

The irreducible representations of the (planar) $D_4$ group are 
(see, e.g., Ref.~\cite{REF:Annett}): 
$\Gamma_1$ or $s$-wave (transforming as $x^2+y^2=1$), 
$\Gamma_2$ [transforming as $xy(x^2-y^2)$], 
$\Gamma_3$ or {$d_{x^2-y^2}$}-wave (transforming as $x^2-y^2$), 
$\Gamma_4$ (transforming as $xy$), and  
$\Gamma_5$ (transforming as the two-component vector $\{x,y\}$). 
(These representations are also commonly denoted  
$A_{1g}, A_{2g}, B_{1g}, B_{2g}$, and $E_{2g}$, respectively.) 
Note that $\Gamma_5$ is a two-dimensional representation, whereas the
other representations are one-dimensional. 

We now focus on the
determination of the terms in the GL free energy 
describing the couplings between
the gradients of $\Gamma_3$ ({$d_{x^2-y^2}$}) and of other  representations.
We consider only the leading
terms of this type, i.e., terms of the form:
\begin{equation}
C_{\mu\nu}\nabla_{\mu}\psi_{\Gamma_3}\nabla_{\nu}\psi_{\Gamma_i},
\label{EQ:grad_gen}\end{equation}
where $\psi_{\Gamma_k}$ is the component of the order
parameter transforming according to representation $\Gamma_k$.
Here, $\mu,\nu=x,y$, and $i=1,\dots,5$.
These terms transform as the (reducible) representation
$\Gamma=\Gamma_3\times\Gamma_i\times\Gamma_5\times\Gamma_5$.
As each term in the free energy must transform as a scalar,
the {\it maximum\/} number of such gradient-coupling terms
is given by the number $N_i$ of times
the identity  ($\Gamma_1$) representation occurs in the decomposition 
of $\Gamma$ into the irreducible representations~\cite{REF:Joynt}. 
$N_i$ is given by the normalized product of characters
corresponding to irreducible representations $\Gamma_3$ and $\Gamma_k$ 
(see, e.g., Ref.~\cite{REF:Landau}).
This gives: $N_i=1$ for $i=1,\dots,4$,
and $N_5=0$. First, we consider the case  $i=2$. A term satisfying
all the symmetries of the group $D_4$ can be written as
\begin{equation}
\frac{C}{2}\big\{\partial_x\psi_{\Gamma_2}\partial_y\psi_{\Gamma_3}^{*}+
\partial_y\psi_{\Gamma_2}\partial_x\psi_{\Gamma_3}^{*}\big\}+\mbox{c.c.},
\label{EQ:grad23}\end{equation}
and, as $N_2=1$, there are no further independent terms~\cite{NOTE:N_2}. 
Next, we consider $i=4$. 
The symmetry $\{x\to y, y\to x\}$ imposes the conditions:
$C_{xx}=-C_{yy}$ and $C_{xy}=C_{yx}=0$, 
while, e.g., the symmetry $\{x\to-x, y\to y\}$ requires that $C_{xx}=0$. 
Therefore, all the
constants are zero and there no are gradient-coupling terms to
leading order for $i=4$. The analysis of cases $i=1,3$ 
has been performed in Ref.~\cite{REF:Joynt}, 
leading to Eqs.~(\ref{EQ:GLd}, \ref{EQ:GLs}),
and $N_5$ is zero. Thus, the only case for which the induction of
the subsidiary component of the order parameter by the current
remains to be considered is that of the coupling between
{$d_{x^2-y^2}$} and $\Gamma_2$-representation [the latter is henceforth 
being referred to as  $d_{xy(x^2-y^2)}$]. 

We denote the component of the order parameter 
corresponding to the {$d_{xy(x^2-y^2)}$}\ -representation by
$a(\bf r)$. In order to construct the GL free energy for the case of
{$d_{xy(x^2-y^2)}$}-{$d_{x^2-y^2}$} coupling, we: (i)~note that the 
structure of terms other
then the mixed gradient terms is the same as for the case of 
the $s$/{$d_{x^2-y^2}$}\ coupling; and (ii)~make use of Eq.~(\ref{EQ:grad23}) 
for the mixed gradient term. The  usual variational procedure then 
leads to the following GL equations for $d$ and $a$:
\begin{mathletters}
\begin{eqnarray}
(-{\gamma_{d}}\nabla^2+{\alpha_{d}})d-{\gamma_{w}}\nabla_x\nabla_y a+
2\sigma_2|d|^2d+\sigma_3|a|^2d
+2\sigma_4a^2d^{*}&=&0,\label{EQ:GLda}\\
(-{\gamma_{a}}\nabla^2+{\alpha_{a}})a-{\gamma_{w}}\nabla_x\nabla_y d+
2\sigma_1|a|^2a+\sigma_3|d|^2a
+2\sigma_4d^2a^{*}&=&0,\label{EQ:GLa}
\end{eqnarray}
\end{mathletters}where 
$\gamma_\rho\equiv\hbar^2/2m_{\rho}$, and $\rho=d, a, w$.  
The current density takes the form
\begin{eqnarray}
{\bf J}&=&\frac{e\hbar}{i{m_{d}}}\big\{d^{*}\nabla d-\mbox{c.c.}\big\}+
\frac{e\hbar}{i{m_{a}}}\big\{a^{*}\nabla a-\mbox{c.c.}\big\}
\nonumber\\
&+&
{\hat x}\frac{e\hbar}{i{m_{w}}}\big\{a^{*}\nabla_yd-\mbox{c.c.}
\big\}
+{\hat y}\frac{e\hbar}{i{m_{w}}}\big\{a^{*}\nabla_x d-\mbox{c.c.}\big\}
\label{EQ:Ja}.
\end{eqnarray}
[As the two last terms in Eq.~(\ref{EQ:Ja}) come from the variation
of the (covariant) mixed gradient terms in the free energy with 
respect to the vector-potential, their structure is different from
that of the analogous terms in Eq.~(\ref{EQ:J})]. As in the case
of $s$/{$d_{x^2-y^2}$}\ -coupling, we assume that the amplitude of $a$ induced 
by the current is small compared to $d$. The first-order perturbative 
calculation analogous to that for the case of the $s$/{$d_{x^2-y^2}$}\ 
-coupling 
leads to the following result for the induced {$d_{xy(x^2-y^2)}$}\ 
-component $a_1$:
\begin{equation}
\frac{a_1}{d_0}=-\frac{\frac{1}{2}{\gamma_{w}} q_0^2 \sin(2\phi)}
{{\gamma_{a}} q^2_0+{\alpha_{a}}+
\frac{5}{2}\,\frac{{\gamma_{w}}^2  q_0^2 \sin^2(2\phi)}{-3{\gamma_{d}} 
q_0^2+{\alpha_{d}}+6\sigma_2d_0^2}
+\frac{{\gamma_{w}}^2}{{\gamma_{d}}}q_0^2+(\sigma_3+2\sigma_4)d_0^2}.
\label{EQ:GLa1}
\end{equation}
We see that in contrast to the case of the $s$/{$d_{x^2-y^2}$}\ -coupling 
[cf.~Eq.~(\ref{EQ:GLs1})], the induced {$d_{xy(x^2-y^2)}$}\-- component is 
zero for
currents flowing along the principal crystallographic axes in the $a-b$
plane (i.e., for  $\phi=0$ or $\phi=\pi/2$) and reaches its maximum
absolute value for currents flowing along the diagonal of the
unit cell (i.~e., for  $\phi=\pi/4$). This difference could be used
in an experiment to determine which of the two couplings
(i.e., $s/{d_{x^2-y^2}}$ or {$d_{xy(x^2-y^2)}$}/{$d_{x^2-y^2}$} ) is realized
in a given HTS material. In the critical region, i.e., 
when $T\to T_{\rm c}$, the term
${\alpha_{a}}$ dominates the denominator of Eq.~(\ref{EQ:GLa1}), 
due to the reasons described in the discussion of the $s$/{$d_{x^2-y^2}$}\ 
-coupling,
and Eq.~(\ref{EQ:GLa1}) then takes the simpler form:
\begin{equation}
\frac{a_1}{d_0}=
-\frac{1}{2}
 \frac{{m_{d}}}{{m_{w}}}
 \frac{|{\alpha_{d}}|}{{\alpha_{a}}}
 (1-f^2)\sin(2\phi).
\label{EQ:GLa1simp}
\end{equation}
As we are not aware of any microscopic theory describing the case
of the {$d_{xy(x^2-y^2)}$}/{$d_{x^2-y^2}$}\ -coupling, we do not know the 
values of the
GL parameters in Eqs.~(\ref{EQ:GLda}, \ref{EQ:GLa}) and, therefore,
cannot give a quantitative estimate for the amplitude of the induced
order parameter. 

\section{Discussion and conclusions}

\begin{figure}
\epsfig{figure=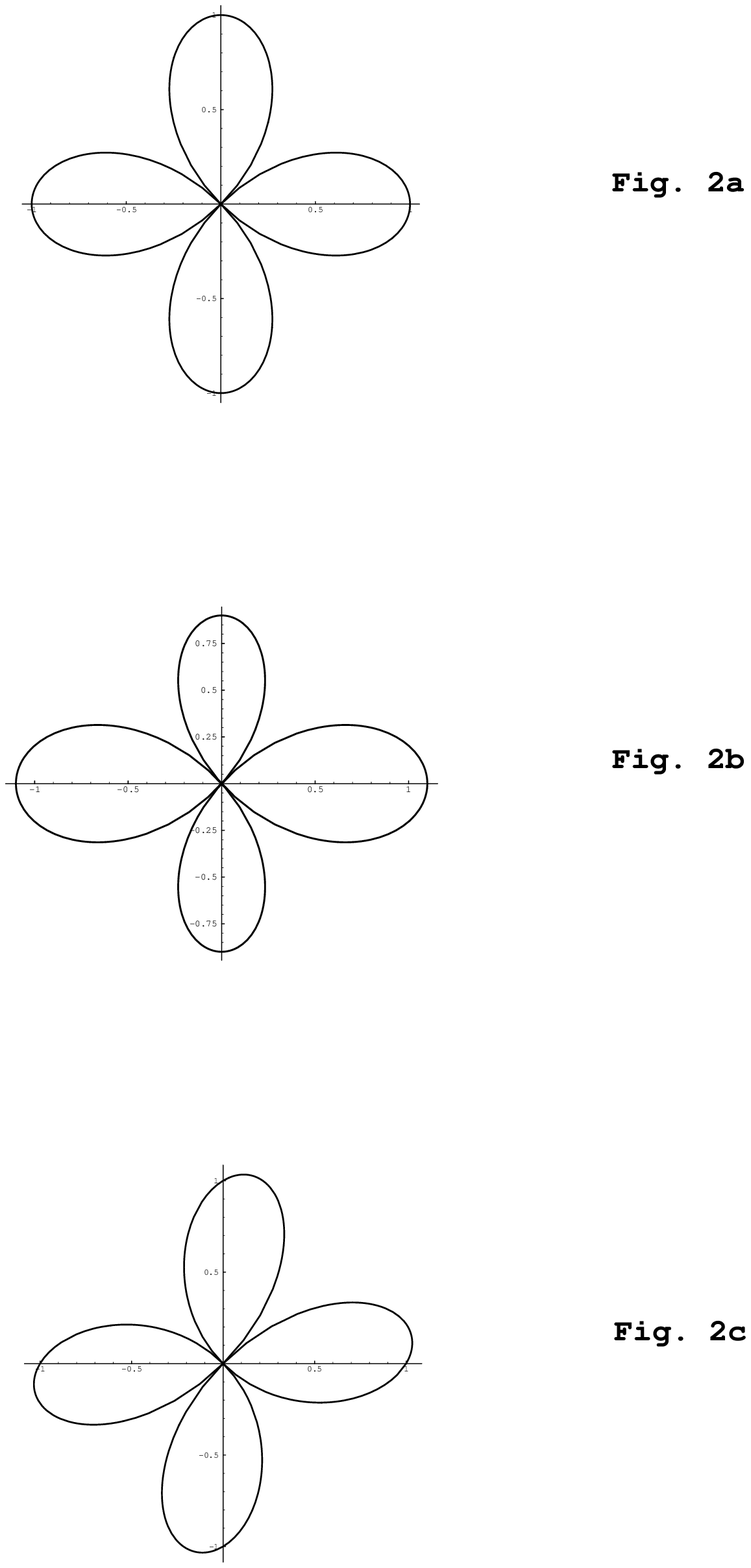, height=18cm,angle=0}
\caption{(a) The $k$-space structure of the superconducting order
parameter with: (a) pure {$d_{x^2-y^2}$} symmetry [$\cos(2 \phi)$]\ ; 
(b) mixed {$d_{x^2-y^2}$} and $s$ symmetry [$\cos(2 \phi) + 0.1$]\ ; 
(c) mixed {$d_{x^2-y^2}$} and {$d_{xy(x^2-y^2)}$} symmetry [$\cos(2 \phi) + 
0.3 \sin (2 \phi)
\cos(2 \phi)$]. Note that for the purpose of illustration, the magnitude of
the $s$-wave component in Fig.~(b) has been chosen to be larger than that
expected from microscopic estimates (see the main text).
\label{FIG:d+s+a}}
\end{figure}

We have seen that the current-induced $s$-wave component introduces an
orthorhombic-like distortion of the $k$-space structure of the order
parameter [Figs.~(\ref{FIG:d+s+a}a,b)]. In contrast, the induced 
{$d_{xy(x^2-y^2)}$}-wave
component distorts the $k$-space structure as indicated in
Fig.~(\ref{FIG:d+s+a}c). Note that the lines of nodes at $\Phi=\pm
\pi/4$ {\it do not\/}
rotate in the {$d_{xy(x^2-y^2)}$}-wave case, and, provided that $a_1 < 2 d_0$, 
no new nodes are introduced. The distortion of the
zero-current, tetragonal structure of Fig.~(\ref{FIG:d+s+a}a) to the
structure of Fig.~(\ref{FIG:d+s+a}b) or Fig.~(\ref{FIG:d+s+a}c) 
(or to a mixture of the latter two) by an externally imposed current
may be experimentally observable using directional probes of the order
parameter. Techniques such as photoemission or tunneling may be
appropriate, provided that sufficiently high resolution can be
obtained.

\acknowledgements
We thank 
G.\ E.\ Blumberg, 
R.\ Giannetta, 
D.\ M.\ Ginsberg, 
N.\ Goldenfeld, 
L.\ H.\ Greene, and 
D.\ J.\ Van Harlingen for useful discussions.  
This work was supported by the NSF under grants
DMR-89-20538 (administered through the Materials Research Laboratory
at the University of Illinois) (MZ and DML) 
and NSF DMR-94-24511 (PMG).



\begin{references}
\bibitem{NOTE:N_2} Note that although the terms
$(\partial_x\psi_{\Gamma_2}\partial_y\psi_{\Gamma_3}^{*} + \mbox{c.c})$ and 
$(\partial_y\psi_{\Gamma_2}\partial_x\psi_{\Gamma_3}^{*} + \mbox{c.c})$ 
transform
separately as scalars, they are related through integration by parts.
\bibitem{REF:Anote}
The four unknown quantities $d_{1}$, $s_{1}$, $q_{1x}$ and $q_{1y}$ 
satisfy four independent equations: 
of these, two are the Ginzburg-Landau equations, (\ref{EQ:GLd},\ref{EQ:GLs}), 
and the remaining two arise from the condition that the perturbations 
do not alter the external current in Eq.~(\ref{EQ:J}).  
To arrive at Eq.~(\ref{EQ:GLs1}), we first eliminate
the quantity $\big(q_{0x}q_{1x}+q_{0y}q_{1y}\big)$ 
from Eqs.~(\ref{EQ:J}) and (\ref{EQ:GLd}) 
to obtain the intermediate result: 
$-3\gamma_{v}\big(q_{0x}^{2}-q_{0y}^{2}\big)s_{1}=
 \big(-3\gamma_{d}q_{0}^{2}+\alpha_{d}+6\beta_{2}d_{0}^{2}\big)d_{1}$.
Then, we eliminate the quantity $\big(q_{0x}q_{1x}-q_{0y}q_{1y}\big)$ 
from Eqs.~(\ref{EQ:J}) and (\ref{EQ:GLs}), 
thus obtaining Eq.~(\ref{EQ:GLs1}).
\end{references}
\end{document}